\documentclass[aps,prl,superscriptaddress,amsmath,amssymb,floatfix, 
twocolumn]{revtex4-2}
\usepackage[T1]{fontenc}
\usepackage[utf8]{inputenc}
\usepackage{amsmath}
\usepackage{graphicx}
\usepackage{babel}
\usepackage{color}

\begin{document}
\title{
Orbital-selective correlation effects
and superconducting pairing symmetry
in a multiorbital $t$-$J$ model for bilayer nickelates
}

\author{Guijing Duan}
\affiliation{Department of Physics and Beijing Key Laboratory of 
Opto-electronic Functional Materials \& Micro-nano Devices,
	Renmin University of China, Beijing 100872, China}

\author{Zhiguang Liao}
\affiliation{Department of Physics and Beijing Key Laboratory of 
Opto-electronic Functional Materials \& Micro-nano Devices,
	Renmin University of China, Beijing 100872, China}

\author{Lei Chen}
\affiliation{Department of Physics \& Astronomy,
	Extreme Quantum Materials Alliance, Smalley Curl Institute,
	Rice University, Houston, Texas 77005,USA}
\affiliation{Department of Physics and Astronomy, Stony Brook University,
	Stony Brook, NY 11794, USA}
	
\author{Yiming Wang}
\affiliation{Department of Physics \& Astronomy,
	Extreme Quantum Materials Alliance, Smalley Curl Institute,
	Rice University, Houston, Texas 77005,USA}

\author{Rong Yu}
\email{rong.yu@ruc.edu.cn}
\affiliation{Department of Physics and Beijing Key Laboratory of 
Opto-electronic Functional Materials \&
	Micro-nano Devices, Renmin University of China, Beijing 100872, China}
\affiliation{Key Laboratory of Quantum State Construction and Manipulation 
(Ministry of Education),
	Renmin University of China, Beijing, 100872, China}

\author{Qimiao Si}
\email{qmsi@rice.edu}
\affiliation{Department of Physics \& Astronomy,
	Extreme Quantum Materials Alliance, Smalley Curl Institute,
	Rice University, Houston, Texas 77005,USA}

\begin{abstract}
The recent discovery of superconductivity in La$_3$Ni$_2$O$_7$
raises key questions about its mechanism and the nature of pairing symmetry.
This system is believed to be described by a bilayer two-orbital Hubbard model. 
The considerations of orbital-selective Mott correlations motivate a bilayer two-orbital $t$-$J$ model and,
accordingly,
we study the superconducting pairing 
in this model.
We obtain an overall phase diagram of 
superconductivity,
where the leading channel 
has either extended 
$s$-wave or $d_{x^2-y^2}$-wave symmetry. 
Our analysis highlights how the
orbital-selective
correlations affect the superconducting pairing 
via the interlayer exchange couplings and
low-energy electronic structure. In particular, we find that 
the dominant orbital for the pairing may change between $z^2$ and $x^2-y^2$ when
 the position of 
the bonding $z^2$ band 
is varied by
 tuning
either the $c$-axis lattice constant or 
electron concentration strength.
We discuss the implications of these results for the 
superconductivity in both bulk 
La$_{3}$Ni$_{2}$O$_{7}$ and 
its thin film counterpart.
\end{abstract}
\maketitle

\begin{figure}
	\includegraphics[width=1\linewidth]{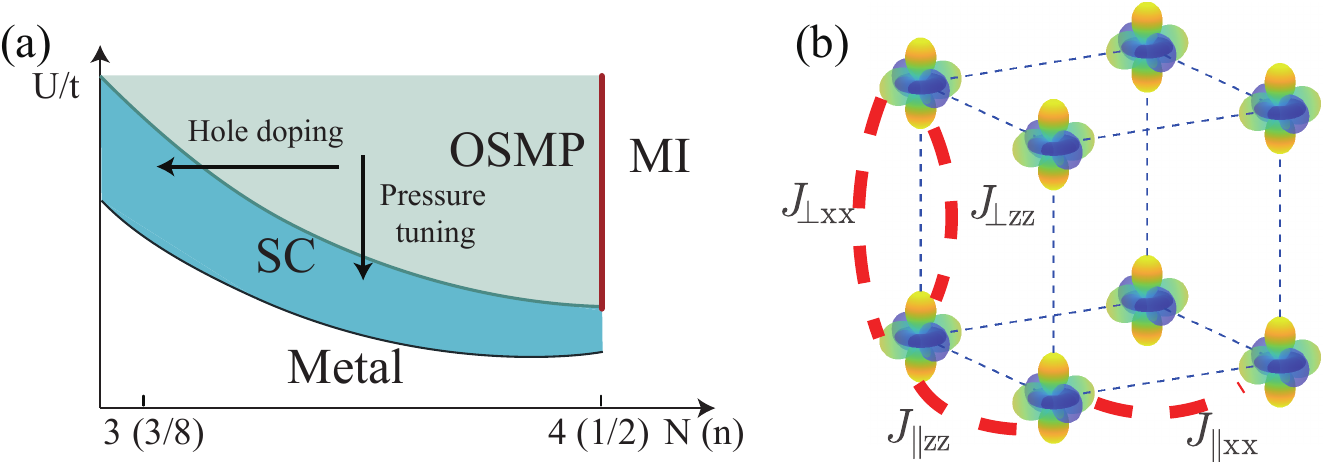}
	\caption{(a) 
	Global phase diagram of orbital-selective correlations, in the parameter 
	space of $U/t$ and electron occupation number at nonzero Hund's coupling
	$J_{\rm{H}}$ of the bilayer two-orbital Hubbard model for 
	La$_3$Ni$_2$O$_7$ \cite{liao2412}. The 
	red line denotes a Mott insulator (MI) where
	both orbitals are localized. The light shaded regime away from
	half-flling stands for an orbital-selective Mott phase (OSMP) where 
	electrons in the $z^2$ orbital are localized while those
	in the $x^2-y^2$ orbital remain itinerant. The system exhibits
	strong orbital-selective behavior near the OSMP and becomes 
	superconducting (SC) at low 
	temperatures. Conceptually, starting from within the OSMP, SC can develop
	by either pressure or hole doing.
	(b) Illustration of the orbital-dependent intralayer and 
	interlayer exchange couplings in the bilayer two-orbital 
	$t$-$J$ model.}\label{fig:1}
\end{figure}

{\it Introduction.~}
The recent discovery of high-temperature superconductivity 
in bilayer nickelate La$_{3}$Ni$_{2}$O$_{7}$ with a transition temperature 
($T_c$) of about $80$ K under high pressure has 
sparked significant research interest \cite{Sun_Nature_2023}. 
Subsequent measurements on this compound and related La$_{2}$PrNi$_{2}$O$_{7}$ 
point to the bulk nature of the superconducting state \cite{Yuan_NP_2024, 
arXiv_2407_05681Cheng} despite possible structural 
imperfections \cite{arXiv_2312_15727Chen,1313,arXiv_2311_12361Sun}. 
In contrast to the infinite-layer nickelate (Sr,\,Nd)NiO$_{2}$ thin
films \cite{LiHwang_Nature_2019}, 
for which the Ni$^{+}$ ion has the same 3d$^{9}$ electron configuration as the high-$T_{\rm c}$ cuprates \cite{Lee_RMP_2006},
La$_{3}$Ni$_{2}$O$_{7}$
has a unique Ni$^{2.5+}$ valence state with a 3d$^{7.5}$
electron configuration.
This configuration implicates multiplicity of the active $3d$ orbitals,
as is the case for the Fe-based superconductors~\cite{YuAbrahamsSi_NatRevMater_2016, Bohmer_NP_2022, SiHussey_PT_2023}.
Together with the restriction to the $z^2$ and $x^2-y^2$
3d orbitals in the low energy sector~\cite{Yao_PRL_2023}, it leads to
a bilayer two-orbital 
Hubbard model with the active electron count of $N=3$ per unit cell.
This encourages the expectation that 
orbital-selective electron correlation effects are important to the electronic 
structure of the system 
\cite{Liao_PRB_2023,ShilenkoLeonov_PRB_2023,LechermannEremin_PRB_2023, 
ZhangDagotto_PRB_2023, 
CaoYang_PRB_2024,OuyangLu_PRB_2024,RyeeWehling_PRL_2024,TianLu_PRB_2024, 
liao2412}. 
These considerations are also anchored on experimental results at ambient pressure:
A recent optical conductivity measurement \cite{LiuWen_NC_2024}
revealed a substantially reduced Drude weight, and  
an angle-resolved photoemission spectroscopy (ARPES) measurement 
\cite{YangZhou_NC_2024} found strongly orbital-dependent enhancement of the 
effective mass in the band structure. 
Taken together, these considerations
suggest that 
La$_{3}$Ni$_{2}$O$_{7}$ is in
 proximity to an orbital-selective Mott phase 
with its 
multiband electronic 
structure strongly renormalized by the interactions.
A global phase diagram [Fig.~\ref{fig:1}(a)]
for the orbital-selective correlations
has recently been advanced \cite{liao2412},
anchored by the half-filling case $N=4$.
The purpose of the present work is to explore the consequence of such correlations for superconductivity
and explore the richness in the pairing states that could arise.

An added motivation comes from the very recent and rapidly
growing experimental reports on superconductivity in thin films of 
bilayer nickelates at ambient pressure \cite{Hwang_Nature_2024, Xue_film2412, 
Bhatt_film2501.08204, 
Hwang_film2501,Chen_film2501}.
In these systems, superconductivity has been observed for films grown on the 
LaSrAlO$_4$ substrate with a compressed average in-plane lattice constant 
compared to the bulk counterpart under ambient pressure, and the observed 
$T_{\rm c}$ varies from 2 K 
to 30 K, which is in general lower than that in the bulk materials. Thin films 
also exhibit distinct structural property and electronic structure from the 
bulk. First, though the in-plane lattice constant is compressed in the film, 
the out-of-plane ($c$-axis) lattice constant is expanded due to the strain 
effect \cite{Hwang_Nature_2024, Bhatt_film2501.08204, Hwang_film2501, 
Xue_film2412}. This 
modifies the interlayer coupling between
the  two Ni 
ions.
Second, a recent 
ARPES measurement~\cite{xuearpes} on the superconducting La$_2$PrNi$_2$O$_7$ thin film 
found a 
sizable hole pocket, likely associated with the $z^2$ bonding band, and the 
entire system is about 0.2 hole per Ni doped from the Ni 3d$^{7.5}$ 
configuration. A hole doping scenario is definitely different from the bulk  
case, which is via pressure tuning.      
These observations raise a number of questions:
What are the key ingredients 
for superconductivity in thin films? 
Does the thin films'
 superconductivity arise in the same way as it does in the bulk materials?
Notwithstanding the extensive theoretical interest in the superconductivity of the bilayer nickelates
\cite{Liao_PRB_2023,QuSu_arXiv_2023, WangHu_arXiv_2024, 
	HeierSavrasov_PRB_2024, ZhanHu_arXiv_2024,
	ChangLi_arXiv_2023,    
	JiangZhang_CPL_2024, HuangZhou_PRB_2023,   
	  XueWang_CPL_2024, ChenLi_PRB_2024, KanekoKuroki_PRB_2024, 
	   SakakibaraKuroki_PRL_2024, 
  JiangKu_PRL_2024, LiuChen_arXiv_2023, YangZhang_arXiv_2024, 
	YangZhang_arXiv_2023, ZhangWeng_PRL_2024, LuYou_arXiv_2023, FanXiang_PRB_2024, 
	ZhengWu_arXiv_2023, SchlomerBohrdt_arXiv_2023, BotzelEremin_arXiv_2024_1, 
	OhZhang_arXiv_2024,hu2025film,yang2025film1,yang2025film2,
	QuSu_PRL_2024, PanWu_arXiv_2023, WangYang_arXiv_2024, QinYang_PRB_2023, 
LuoYao_npjQM_2024, YangZhang_PRB_2023, LuWu_PRL_2024, LuWu_PRB_2024, 
KakoiKuroki_PRB_2024, MaWu_arXiv_2024, YangWang_PRB_2023, ZhangDagotto_NC_2024},
the issues as outlined above regarding superconductivity have yet to be analyzed theoretically.

In this Letter, we do so by
studying the superconducting pairing of the bilayer nickelates 
with the particular attention on the orbital-selective electronic correlations. 
Anchored by the global phase diagram in Fig.~\ref{fig:1}(a), 
which characterizes the strong 
orbital-selective behavior
through a proximity to a 
two-orbital Mott insulator \cite{liao2412}, 
superconductivity 
can be understood as arising
by either pressure tuning (which reduces the correlation strength)
or carrier doping. The proximate Mott 
nature provides the basis for the emergence of
an effective bilayer two-orbital $t$-$J$ model,
which we use to study
the superconductivity.
We show that the dominant superconducting pairing state
has either an extensive $s$-wave or a 
$d_{x^2-y^2}$-wave symmetry with the gap anisotropy influenced by the 
interlayer exchange couplings. Moreover, we find that tuning the position of 
the $z^2$ bonding band by either strain effect or hole doping can induce an orbital-selective pairing crossing, with
the dominant orbital contribution to the pairing channel changing  from 
$z^2$ to $x^2-y^2$. These 
results shed light on the nature of the superconducting states in both bulk 
bilayer nickelates and related thin films.

{\it Model and method.~}
To consider the electron correlation effects in 
La$_3$Ni$_2$O$_7$, we start from a bilayer two-orbital Hubbard model 
\cite{liao2412}. The slave-spin calculation on this model suggests that the 
system with $N=3$ electrons per unit cell is in proximity to a putative Mott 
insulator at $N=4$ 
and exhibits strong orbital-selective Mott behavior, as illustrated in 
Fig.~\ref{fig:1}(a). The low-energy effective model capturing the strong 
correlation effects is then a bilayer two-orbital $t$-$J$ model 
\cite{Liao_PRB_2023}, with the Hamiltonian 
that
reads
\begin{align}
\mathcal{H} & =
\sum_{i\delta\alpha\beta\sigma}\frac{\sqrt{Z_{\alpha}Z_{\beta}}}{2} 
t_{\delta}^{\alpha\beta}f_{i\alpha\sigma}^{\dagger}f_{i+\delta\beta\sigma} 
+\sum_{i\alpha\sigma}(\epsilon^\prime_{\alpha}-\mu) 
f_{i\alpha\sigma}^{\dagger}f_{i\alpha\sigma}\nonumber \\
 & 
 +\sum_{i\delta\alpha} J_{\delta\alpha\alpha} 
 (\boldsymbol{S}_{i\alpha}\cdot\boldsymbol{S}_{i+\delta\alpha} 
 -\frac{1}{4}n_{i\alpha}n_{i+\delta\alpha})\label{hamiltonian}
\end{align}
Here we have employed the slave-spin method 
\cite{Yu_PRB_2012,Yu_PRB_2017,LanataHellsing_PRB_2012} to renormalize the 
kinetic part of 
the Hamiltonian, with $Z_{\alpha}$ the quasiparticle spectral weight of the 
$\alpha$ orbital. We take the $Z_{\alpha}$ values of corresponding 
$U$ and Hund's coupling $J_{\rm{H}}$ from the calculation in 
Ref.~\cite{liao2412}. $f_{i\alpha\sigma}^{\dagger}$ creates a spinon at site 
$i$, 
in orbital $\alpha$, and spin projection $\sigma$. $\epsilon'_{\alpha}$ refers 
to the renormalized energy level
of orbital $\alpha$, $\mu$
is the chemical potential, and $t_{\delta}^{\alpha\beta}$ the hopping
matrix along the $\delta\ (=\hat{x},\hat{y},\hat{z})$ direction of the 
tight-binding model. The 
orbital 
index $\alpha=z,\ x$ correspond 
to the $z^{2}$ and $x^{2}-y^{2}$ orbitals, respectively. 
The spin operator of the local moment is defined as 
$\text{\ensuremath{\boldsymbol{S}_{i\alpha}}}=\frac{1}{2}\sum_{ss^\prime} 
f_{i\alpha
 s}^{\dagger}\boldsymbol{\sigma_{ss^\prime}}f_{i\alpha s^\prime}$,
where $\boldsymbol{\sigma}$ represents the Pauli matrices.
This approach is a generalization of the slave-boson theory 
\cite{Lee_RMP_2006,Liao_PRB_2023,kotliarslaveboson} to the finite $U$ 
and multiorbital case. 

The effects of electron correlations have been taken into account by 
the renormalized electronic structure and the orbital-dependent intralayer and 
interlayer exchange 
interactions $J_{\parallel\alpha\alpha}$ and $J_{\perp\alpha\alpha}$. Here we 
keep only the leading 
exchange interactions as illustrated in Fig.~\ref{fig:1}(b).
We assume 
the intralayer exchange couplings take the 
strong-coupling form $J_{\parallel 
\alpha\alpha}= 4t^{2}_{\parallel\alpha\alpha}/U$. 
It is expected that the interlayer exchange couplings are
sizable.
 To examine their effects, we treat them over a parameter range
in the model. Note that here we also include an effective interlayer coupling 
in the $x^{2}-y^{2}$ orbital, $J_{\perp xx}$, which 
can be mediated through the strong Hund's coupling 
\cite{LuWu_PRL_2024}. 

The superconducting pairing in the model of Eq.\eqref{hamiltonian}
is studied by a Bogoliubov Hubbard-Stratonovich
decomposition of the
exchange interactions in the spin singlet sector: 
\begin{eqnarray}
 && J_{\delta\alpha\alpha}
 \left(\boldsymbol{S}_{i\alpha}\cdot\boldsymbol{S}_{i+\delta\alpha}
  -\frac{1}{4}n_{i\alpha}n_{i+\delta\alpha}\right) \nonumber\\
  &&\approx 
 -\frac{1}{2}\left(\hat{\Delta}_{\delta \alpha}^{\dagger} 
 \hat{\Delta}_{\delta\alpha} + \rm{h.c.} - |\Delta_{\delta\alpha}|^2\right),
\end{eqnarray}
where $\hat{\Delta}_{\delta\alpha} 
=f_{i\alpha\downarrow}f_{i+\delta\alpha\uparrow} 
-f_{i\alpha\uparrow}f_{i+\delta\downarrow}$, and the gap function 
$\Delta_{\delta\alpha}=\langle 
\hat{\Delta}_{\delta\alpha} \rangle$ is solved in a self-consistent way 
\cite{yunc2013}. The superpositions of these gap functions then transform 
according 
to different irreducible representations of the $D_{4h}$ group, with either 
$s$- or $d$-wave symmetry \cite{Liao_PRB_2023}. A complete symmetry 
classification is presented in the Supplemental Materials (SM) \cite{SM}. 
 
\begin{figure}
	\includegraphics[width=1\linewidth]{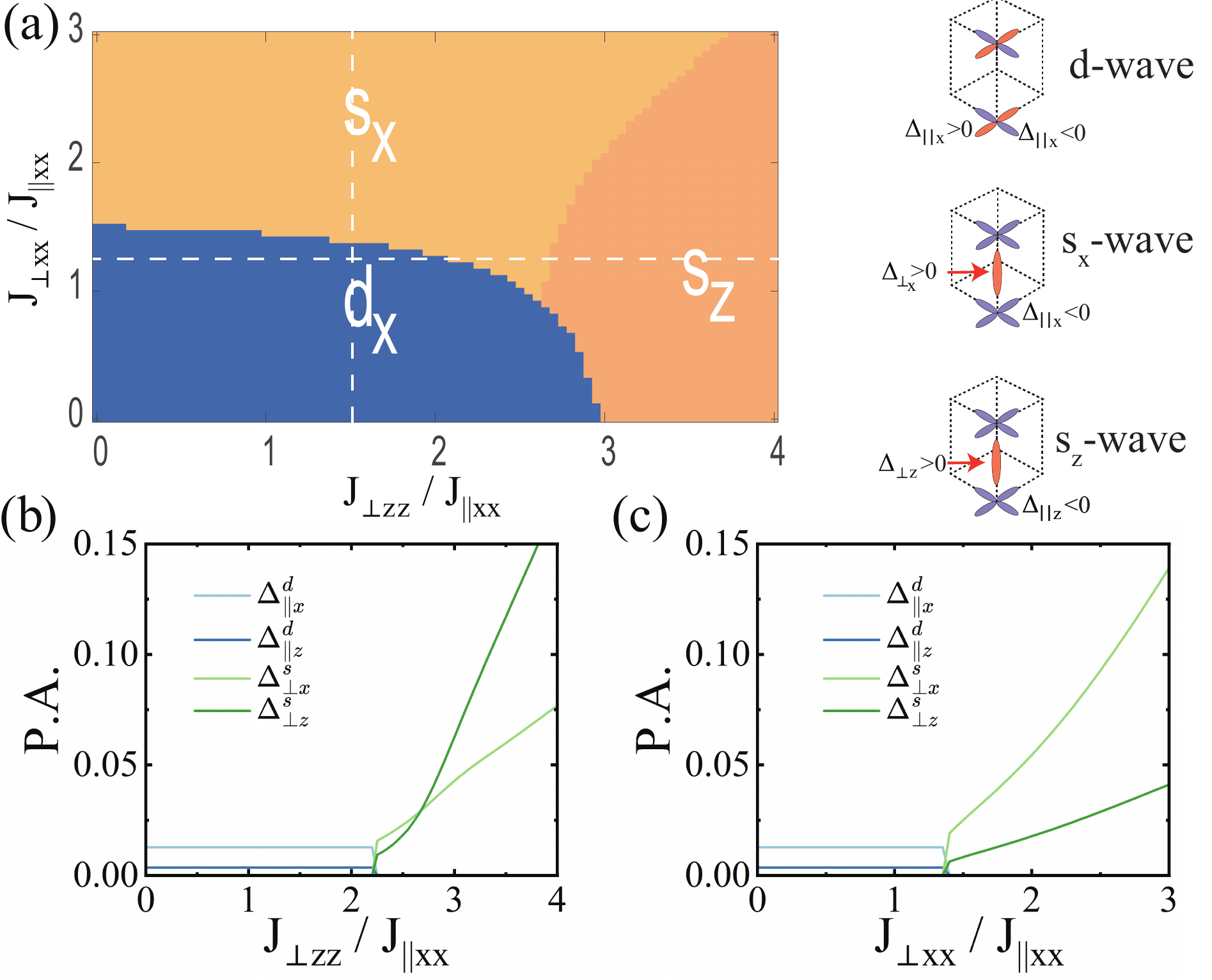}
	\caption{(a) Superconducting phase diagram with $J_{\perp xx}$ and 
	$J_{\perp zz}$ at $U=4.8$ eV, which gives 
	$J_{\parallel xx}/W=0.05$ and 
	$J_{\parallel zz}/W=0.0125$, where $W$ is the bandwidth of the 
	tight-banding 
	part of the bilayer two-orbital model. 
	Here $d_x$, $s_x$, and $s_z$ refer to distinct 
	leading pairing symmetries. d$_{x}$ denotes the $d_{x^2-y^2}$-wave $B_{1g}$ 
	symmetry with in-plane pairing in the $x^2-y^2$ orbital, while s$_{x(z)}$
	represents the extensive $s$-wave $A_{1g}$ symmetry with interlayer pairing 
	in the $x^2-y^2$ ($z^2$) orbital. Real-space pairing structures of these 
	three channels are illustrated in the right column of panel (a). (b) 
	Evolution of dominant pairing amplitudes (P.A.) along the horizontal dashed 
	line $J_{\perp xx}/J_{\parallel xx}=1.2$ in panel (a), showing transition 
	and crossover of the pairing symmetry with the interlayer exchange coupling 
	$J_{\perp zz}$. (c) Same as (b) but along $J_{\perp zz}/J_{\parallel 
	xx}=1.5$.}\label{fig:2}  
\end{figure}


{\it Superconducting pairing symmetry.~}
We study how the superconducting pairing amplitude evolves with the interlayer 
exchange couplings $J_{\perp xx}$ and $J_{\perp zz}$, and a typical phase 
diagram for $U=4.8$ eV and $J_{\rm{H}}/U=0.2$ is shown in 
Fig.~\ref{fig:2}(a). For other $U$ and $J_{\rm{H}}$ values generating the strong 
correlation effects, the phase diagram looks similarly. 
In this phase diagram the leading pairing channel has either a 
$d_{x^2-y^2}$-wave $B_{1g}$ or an extensive 
$s$-wave $A_{1g}$ symmetry. The $d$-wave channel is dominant by the in-plane 
pairing of electrons in the $x^2-y^2$ orbital, which is stabilized in the weak 
interlayer coupling regime of the phase diagram. Pairing amplitudes of several 
leading pairing channels are shown in 
Fig.~\ref{fig:2}(b) and (c), and those in all channels are shown in Fig.~S1 of 
the SM \cite{SM}. From these 
results  we see that
increasing either $J_{\perp zz}$ or $J_{\perp 
xx}$, the pairing symmetry transitions to $s$-wave where the dominant gap 
function is the interlayer pairing. It can be either $\Delta_{\perp z}^s$ within 
the 
$z^2$ orbital or 
$\Delta_{\perp x}^s$ within the $x^2-y^2$ orbital, depending on the strength of 
$J_{\perp xx}$ and $J_{\perp zz}$. These two regimes, both with the $s$-wave 
symmetry, are separated by a crossover as shown in the phase diagram of 
Fig.~\ref{fig:2}(a). In a broad regime of the $s$-wave pairing phase 
we find the in-plane gap function has opposite sign to the interlayer one in 
the leading pairing orbital, as illustrated in the right column of 
Fig.~\ref{fig:2}(a). Such a sign structure helps stabilize the superconductivity 
by avoiding nodes on the Fermi surface (see below).    
The evolution
of pairing symmetry highlights the sensitivity of the superconducting
state to the relative strengths of interlayer exchange interactions.
Notably, along the boundaries between the $s$-wave and $d$-wave
phases, a mixed $s+id$ pairing symmetry may emerge \cite{yunc2013,hu2018}.
However, 
this state is
found to be
 nearly degenerate in energy
with the pure $d$-wave phase
and thus
is not explicitly marked in the phase diagram.
\begin{figure}
\includegraphics[width=1\linewidth]{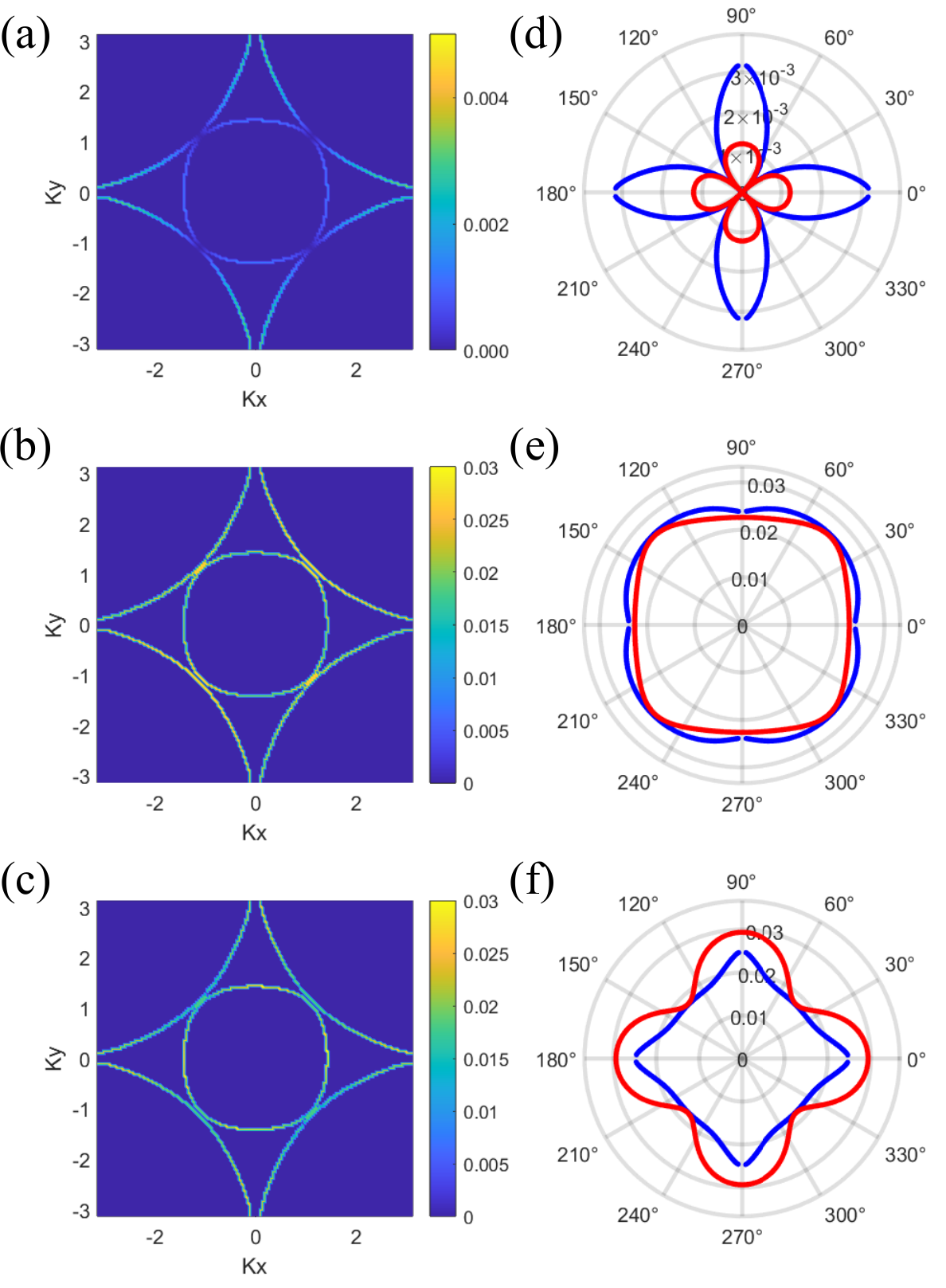}
\caption{Superconducting gaps projected onto the Fermi
surface in phases with different pairing symmetries and corresponding angular 
dependence of the gaps: (a) and (d) for $d$-wave pairing at $J_{\perp 
xx}/J_{\parallel xx}=0.5$ and $J_{\perp zz}/J_{\parallel xx}=2.5$; (b) and (e) 
for $s_{x}$-wave pairing at $J_{\perp xx}/J_{\parallel xx}=2.5$ and $J_{\perp 
zz}/J_{\parallel xx}=2.5$; (c) and (f) for $s_{z}$-wave pairing at $J_{\perp 
xx}/J_{\parallel xx}=1$ and $J_{\perp zz}/J_{\parallel xx}=4$.}\label{fig:3}
\end{figure}

We next
investigate the distribution of the superconducting gap
on the Fermi surface. 
Representative results at specific parameter sets corresponding to the $d_x$, 
$s_x$, and $s_z$ pairing are shown in Fig.~\ref{fig:3}. 
For the $d$-wave pairing, the gap function, taking the form of 
$\Delta_{\parallel}^d (k)\propto\cos(k_{x})-\cos(k_{y})$, changes sign across 
$k_x=\pm k_y$ and    
features nodes along these two directions. This causes strongly anisotropic 
superconducting gap. 
In contrast, the $s$-wave phase is fully gapped and nodeless 
because the dominant interlayer pairing $\Delta_{\perp}^s$ 
enforces a uniform gap structure without any sign change. 
However, strong competition between pairing in the two orbitals as well as the 
interplay between the interlayer pairing $\Delta_{\perp}^s$ and the intralayer 
pairing $\Delta_{\parallel}^s \propto\cos(k_{x})+\cos(k_{y})$ 
cause a complex gap structure that cannot be described by a single gap function 
and may give rise to ansiotropic superconducting gap. 
These are reflected in the primarily distinct gap anisotropies in the $s_{x}$ 
and $s_{z}$ regimes of the phase diagram.  
In the $s_{x}$ regime, the gap is almost isotropic with minima 
along the $\pm k_x/k_y$ direction. But in the $s_{z}$ regime, due to the 
influence of the subleading pairing in the $x^2-y^2$ orbital, the gaps on the 
Fermi surfaces can exhibit substantial anisotropy with gap minima along the 
$k_x=\pm k_y$ direction.  

\begin{figure}
\includegraphics[width=1\linewidth]{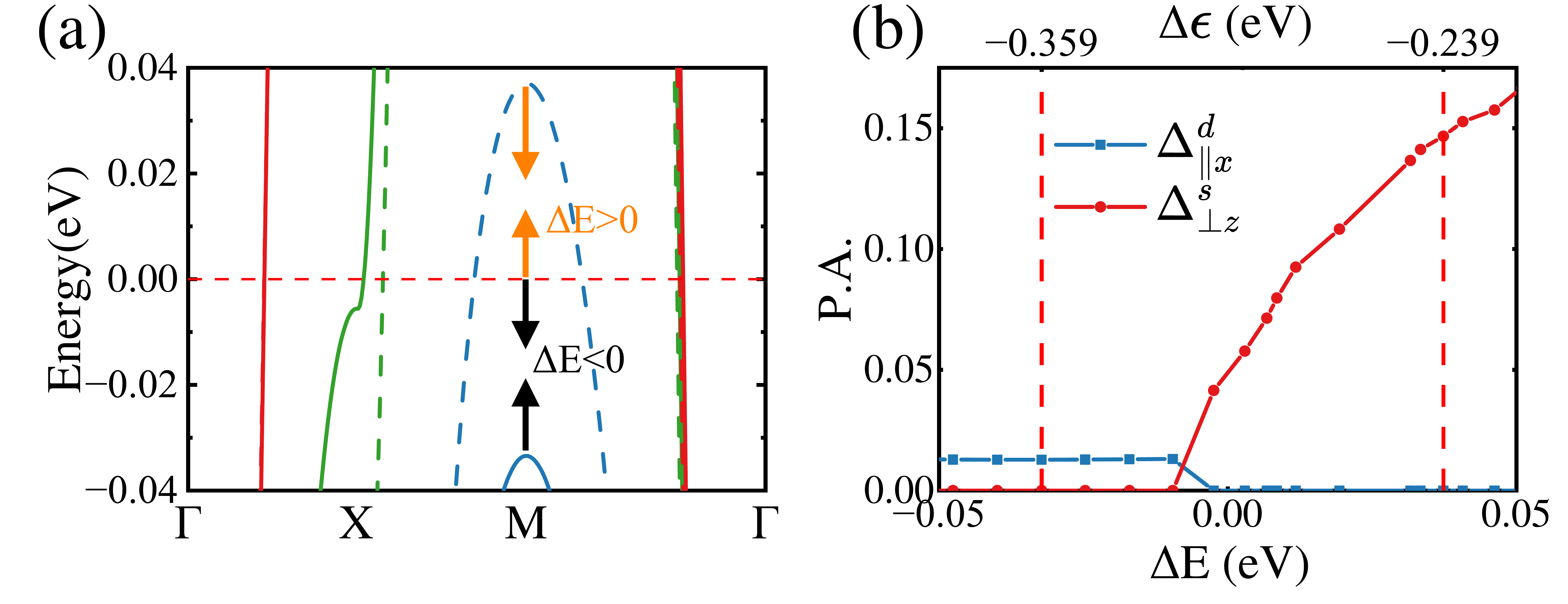}
\caption{
(a) Band structures with different crystal splitting $\Delta\epsilon=-0.239$ 
	eV (dashed line) and $\Delta\epsilon=-0.359$ eV (solid line), respectively. 
$\Delta E$ denotes the 
energy of the $z^2$ bonding band top to the Fermi level, which is sensitive to 
$\Delta\epsilon$. 
(b) Leading pairing amplitudes (P.A.) as a function of $\Delta E$ and 
$\Delta\epsilon$ showing a transition between $d$-wave and $s$-wave pairing 
symmetry. The exchange couplings are taken to be $J_{\perp 
zz}/J_{\parallel 
xx}=2.0$ and 
$J_{\perp xx}/J_{\parallel xx}=0.5$.
}\label{fig:4}
\end{figure}

{\it Pairing crossing between the 
	$z^2$ and $x^2-y^2$ 
	orbitals
	induced by the shift of 
the $z^2$ bonding band.~}
We now
 investigate the evolution
of superconducting pairing symmetry as a function of the top position
of 
the $z^2$ orbital bonding band (near the M point of the Brillouin zone);
$\Delta E$ marks the separation of this band top
to the Fermi level.
 We tune 
$\Delta E$ by varying the crystal splitting between the $x^2-y^2$ and $z^2$ 
orbitals, $\Delta\epsilon=\epsilon^\prime_z - \epsilon^\prime_x$, in the 
tight-binding part of the Hamiltonian. As shown in Fig.~\ref{fig:4} for 
$J_{\perp zz}/J_{\parallel xx}=2.0$ and $J_{\perp xx}/J_{\parallel xx}=0.5$, 
when $\Delta E>0$, where the $z^2$ bonding band crosses the Fermi level, we find 
the leading pairing symmetry is always $s_z$. Decreasing $\Delta E$ to negative 
value, where the $z^2$ bonding band sinks to below the Fermi level, the leading 
pairing symmetry transitions to $d_x$. Comparing to the $s_z$ pairing, the 
pairing amplitude of the $d_x$ channel is largely reduced and becomes almost 
independent of $\Delta E$ away from the transition regime. In other regimes of 
the phase diagram, a similar crossover between the $s_z$ and $s_x$ pairing 
induced by $\Delta E$ can take place, as shown in Fig.~\ref{fig:SM2} of the SM 
\cite{SM}. 
Such a crossover/transition between the $z^2$ and $x^2-y^2$ orital 
pairing 
can be understood as follows: 
The system contains singlet pairing of electrons in the $z^2$ orbital between 
the top and bottom layers. Once the $z^2$ orbital bonding band crosses the Fermi 
level, the coherent movement of these prepaired 
electrons become superconducting with the $s$-wave symmetry. On the other hand, 
when the $z^2$ orbital bonding band is below the Fermi level, though the 
electrons could be locally paired, a global coherence is not established within 
the 
$z^2$ orbital, given its proximity to the Mott localization. 
In this limit, however, the superconductivity can be carried by electrons in the 
$x^2-y^2$ orbital, with a $d$-wave (or an $s$-wave) symmetry. Given the 
in-plane pairing structure and the much weaker interlayer hopping in the 
$x^2-y^2$ orbital, the pairing amplitude of the $d_x$ pairing is almost 
independent of $\Delta E$ and largely reduced compared to the $s_z$-wave case.   

{\it Discussion and conclusion.~}
In our calculations, the strong 
orbital-selective correlations affect the superconducting pairing via both 
renormalizing the non-interactingelectronic structure and inducing 
orbital-dependent exchange interactions. As a result, the leading pairing 
channel is generally orbital dependent, dominated by either $z^2$ or $x^2-y^2$ 
orbital, as shown in the phase diagram of 
Fig.~\ref{fig:2}(a). In particular, delocalization of electrons in the $z^2$ 
orbital from the OSMP (Fig.~\ref{fig:1}(a))
plays a crucial role in stabilizing the 
$s$-wave pairing. This orbital may directly participate in superconductivity, 
causing the $s_z$ pairing in the phase diagram. In this regime the itinerancy of 
electrons in this orbital helps establish a global phase coherence of the 
superconductivity. Even in the $s_x$ regime where the dominant pairing 
originates from the $x^2-y^2$ orbital, the $z^2$ orbital helps
mediate pairing
by enhancing $J_{\perp xx}$ through Hund's coupling.  

Our results also show that the ratio of interlayer to intralayer exchange 
coupling is important for determining the superconducting state in 
La$_3$Ni$_2$O$_7$. Based on a previous slave-spin calculation \cite{liao2412} 
we roughly estimate
 $J_{\perp zz}/J_{\parallel xx}\sim 1$-$3$ and $J_{\perp 
xx} < J_{\perp zz}$. Correspondingly, the system is in a regime of the phase 
diagram where $d_x$, $s_x$, and $s_z$ 
pairing channels are in strong competition. Note that although our estimated 
exchange couplings are within the same order as those 
suggested by
 resonant 
inelastic 
X-ray scattering and inelastic neutron scattering measurements at ambient 
pressure \cite{xie2024strong,fengrixs}, 
precise determination of the values awaits future experimental and theoretical studies.
 Examining the gap anisotropy would 
be 
very helpful to determine the pairing symmetry. However, as 
shown in Fig.~\ref{fig:3}, both the $d_x$ and $s_z$ pairing can be strongly 
anisotropic and the gap minima of the two are located along the same direction 
of the Brillouin zone. Therefore, phase sensitive measurements would be 
necessary to distinguish these two pairing symmetries.

As shown in Fig.~\ref{fig:4}, the pairing symmetry can be sensitive to the 
position 
of the $z^2$ bonding band, which can be tuned by either carrier doping or 
crystal field splitting $\Delta \epsilon$. 
Note that the evolution of superconductivity we find here crucially depends on 
the change in the Fermi surface and, thus, is different from what happens to the 
effect of crystal field splitting in certain weak-coupling context 
\cite{XiaChen_NC_2025}. In our theory, increasing $\Delta \epsilon$ raises 
the onsite energy of the $z^2$ orbital, corresponding to hole doping the $z^2$ 
bonding band while electron doping the $x^2-y^2$ orbital. In real materials, 
$\Delta \epsilon$ is closely related to the out-of-plane lattice constant $c$.  
Increasing $c$ elongates the outer Ni-O bond distances and reduces the onsite 
energy of the $z^2$ orbital, $\epsilon_{z}$.
As already mentioned,
this may trigger a transition of 
the pairing symmetry from $s_z$ 
to $d_x$ by pushing the $z^2$ bonding band to be below the Fermi level. Note 
that the pairing amplitude of the $d$-wave channel is largely reduced. This may 
explain why 
$T_{\rm{c}}$ is significantly reduced
 in La$_3$Ni$_2$O$_7$ thin films \cite{Hwang_Nature_2024}: In the 
films, $c$ is elongated compared to the bulk counterpart as a strain effect by 
compressing the in-plane lattice constants;
according to our results, the 
pairing in these thin films could be $d$-wave, in contrast to the likely 
$s$-wave in the bulk.  
Following this idea, we examine the pressure effects in the bulk. The 
hydrostatic pressure would compress both in-plane and out-of-plane lattice 
constants. With $c$ compressed, the $z^2$ bonding band is raised across the 
Fermi level, which is likely to stabilize the $s$-wave pairing with a high 
$T_{\rm{c}}$.
As for the recently observed superconductivity in La$_2$PrNi$_2$O$_7$ thin 
films \cite{Hwang_film2501, Xue_film2412}, an ARPES study 
implicates an effective 
hole doping as playing a crucial role 
in stabilizing the $T_{\rm{c}}$ up to about 30 
K. It is possible that in this material, there is strong competition between 
effects of the crystal field splitting and hole doping,
such that the pairing symmetry 
stays the same as in the bulk 
system under pressure.  

In conclusion, based on the strong orbital-selective correlations and proximity 
to a Mott insulator, we 
consider an effective bilayer two-orbital $t$-$J$ 
model to study the superconducting pairing of La$_3$Ni$_2$O$_7$. We present a 
phase diagram 
with respect to the
 interlayer exchange couplings and find the leading pairing 
to have either an extensive $s$-wave or a 
$d_{x^2-y^2}$-wave symmetry. We show that tuning the position of 
the $z^2$ bonding band by either strain effect or carrier doping can 
induce a transition between 
$z^2$ and $x^2-y^2$ orbital contribution to the leading pairing 
channel. These 
results help elucidate the puzzling superconducting states in both 
the
bulk 
bilayer nickelates and 
their thin film counterparts.

\begin{acknowledgments}
	We thank W. Ding, H. Y. Hwang, W. Ku, W. Li, K.-S. Lin, X. Lu, L. Sun, F. 
	Wang, M. Wang, C. 
	Wu, F. Yang, Y.-f. Yang, and G.-M. Zhang
	for useful discussions. This work has in part been supported by
	the National Science Foundation of China (Grants 12334008 and 12174441).
	Work at Rice was primarily supported
	by the U.S. Department of Energy, Office of Science, Basic Energy Sciences,
	under Award No. DE-SC0018197, and by
	the Robert A.\ Welch Foundation Grant No.\ C-1411.
	Q.S. acknowledges the hospitality of the Aspen Center for Physics,
	which is supported by the NSF grant No. PHY-2210452.
\end{acknowledgments}
 \bibliographystyle{apsrev}

\bibliography{La3Ni2O7_Pairing}

\pagebreak


\onecolumngrid
\newpage
\newcounter{equationSM}
\newcounter{figureSM}
\newcounter{tableSM}
\stepcounter{equationSM}
\setcounter{equation}{0}
\setcounter{figure}{0}
\setcounter{table}{0}

\makeatletter
\renewcommand{\theequation}{\textsc{S}\arabic{equation}}
\renewcommand{\thefigure}{\textsc{S}\arabic{figure}}
\renewcommand{\thetable}{\textsc{S}\arabic{table}}
\begin{center}
	{\large{\bf SUPPLEMENTAL MATERIAL -- 
Orbital-selective correlation effects and superconducting pairing symmetry
in a multiorbital $t$-$J$ model for bilayer nickelates }}
\end{center}
\setcounter{page}{1}
\begin{center}
	Guijing Duan$^1$, Zhiguang Liao$^1$, Lei Chen$^{2,3}$, Yiming Wang$^2$, 
	Rong Yu$^{1,4}$, Qimiao Si$^2$\\
	\quad\\
	$^1$\textit{School of Physics and Beijing Key Laboratory of Opto-electronic
		Functional Materials and Micro-nano Devices, Renmin University of China,
		Beijing 100872, China}\\
	$^2$\textit{Department of Physics \& Astronomy,
		Extreme Quantum Materials Alliance, Smalley Curl Institute,
		Rice University, Houston, Texas 77005,USA}\\	
	$^3$\textit{Department of Physics and Astronomy, Stony Brook University,
		Stony Brook, NY 11794, USA}\\
	$^4$\textit{Key Laboratory of Quantum State Construction and Manipulation
		(Ministry of Education), Renmin University of China, Beijing, 100872, 
		China}
\end{center}

\section{Symmetry classification of superconducting gap functions}
Here we present a symmetry classification of the superconducting gap functions 
of the bilayer two-orbital $t$-$J$ model by analyzing 
their transformation properties under the tetragonal $D_{4h}$ group. We consider 
only the orbital diagonal exchange couplings in the model, and the gap functions 
are also orbital diagonal in this work.
The resulting symmetry assignments in each orbital are summarized in 
Table~\ref{Tab:S1}.
Basically, the in-plane $s$-wave and $d$-wave pairing channels are formed by 
linear combinations of the gap functions in the real space, {\it i.e.}, 
$\Delta_{\parallel}^{s/d} = \Delta_{\hat{x}}\pm \Delta_{\hat{y}}$. These 
in-plane $s$ and $d$ pairing channels have $\cos k_x \pm \cos k_y$ form factors, 
respectively. These pairing 
channels in the two layers are then combined to two, transforming symmetrically 
and antisymmetrically under the inversion symmetry (distinguished by the Pauli 
matrices $\eta_0$ and $\eta_3$, respectively), respectively. For the 
interlayer pairing, $\Delta_{\perp}^{s} = \Delta_{\hat{z}}$ is uniformly 
$s$-wave, 
given that
the dominant
 interlayer exchange coupling corresponds to the nearest neighbor term
$J_{\perp}$.

\begin{table}[!h]
	\centering
	\begin{tabular}{cccc}
		\hline\hline
		$\Delta(\mathbf{r})$ & $\Delta(\mathbf{k})$ & Symmetry \\
		\hline
		$(\Delta_{\hat{x}} + \Delta_{\hat{y}})\eta_0$ & $s_{x^2+y^2} \eta_0$ & 
		$A^{1g}$ \\
		$(\Delta_{\hat{x}} - \Delta_{\hat{y}})\eta_0$ & $d_{x^2-y^2} \eta_0$ & 
		$B^{1g}$ \\
		$(\Delta_{\hat{x}} + \Delta_{\hat{y}})\eta_3$ & $s_{x^2+y^2} \eta_3$ & 
		$A^{2u}$ \\
		$(\Delta_{\hat{x}} - \Delta_{\hat{y}})\eta_3$ & $d_{x^2-y^2} \eta_3$ & 
		$B^{2u}$ \\
		$\Delta_{\hat{z}}$ & $s_z $ & $A^{1g}$ \\
		\hline\hline
	\end{tabular}
	\caption{Superconducting pairing symmetry of the effective multiorbital 
	$t$-$J$ model.
		$\Delta(\mathbf{r})$ and $\Delta(\mathbf{k})$ refer to 
		the gap functions in 
		real and momentum space, respectively.
		Their symmetry is characterized by the corresponding irreducible 
		representation of the tetragonal $D_{4h}$ group.
		$\eta_0$ and $\eta_3$ refer to the $2\times2$ unit marix and 
		$z$-component of the Pauli 
		matrices,  respectively.
		$\hat{x}(\hat{y},\hat{z})$ refers to 
		the unit vector along the $x(y,z)$ 
		direction.}\label{Tab:S1}
\end{table}

\begin{figure}[!h]
	\includegraphics[width=0.7\linewidth]{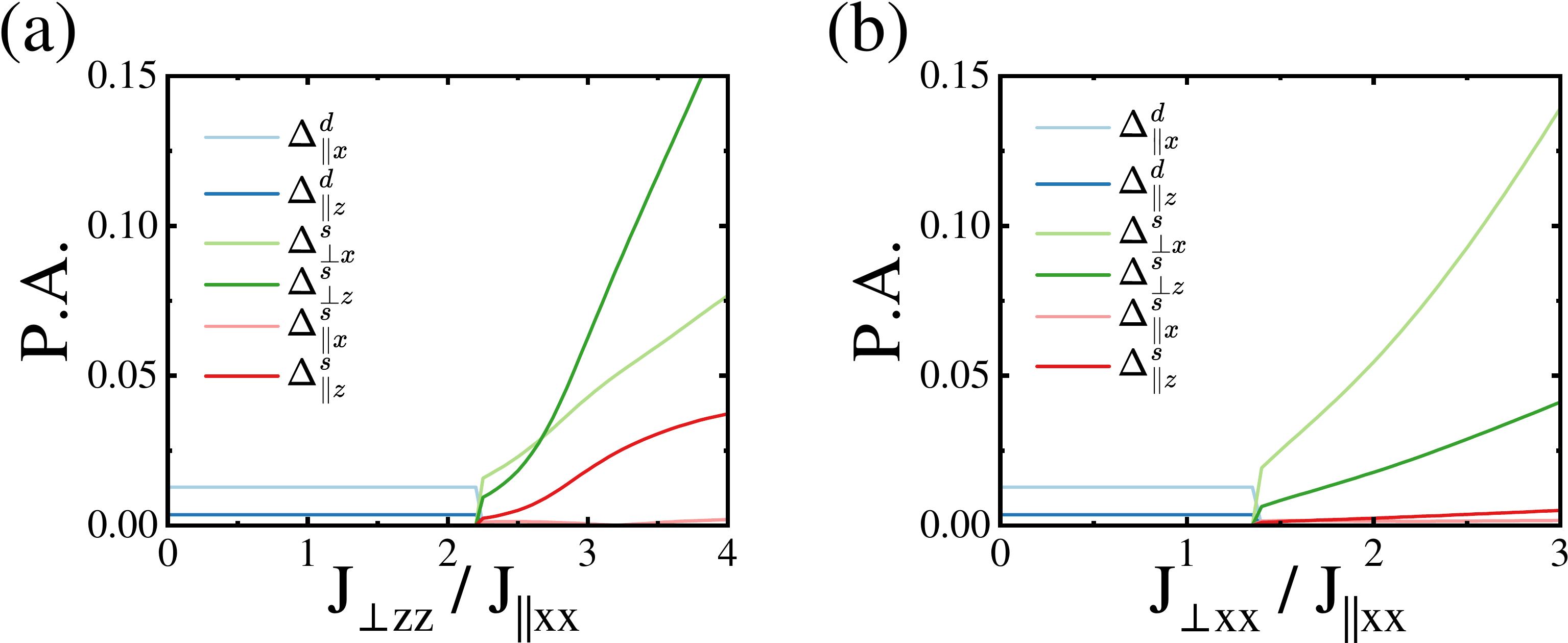}
	\caption{(a) Evolution of all pairing amplitudes (P.A.) along the 
	horizontal dashed
		line $J_{\perp xx}/J_{\parallel xx}=1.2$ in Fig.~2(a). (b) 
		Same as (a) but along $J_{\perp zz}/J_{\parallel
			xx}=1.5$.
	}\label{fig:SM1}
\end{figure}

Pairing amplitudes along dashed lines corresponding to model parameters in 
Fig.~\ref{fig:2} are shown in Fig.~\ref{fig:SM1}. In the calculation, we find 
only three pairing channels in each orbital, $\Delta_{\perp x(z)}^s$, 
$\Delta_{\parallel x(z)}^s$ and $\Delta_{\parallel x(z)}^d$,
with $A^{1g}$ and $B^{1g}$ symmetries, respectively. The paring 
amplitudes of the leading pairing channels have been shown in 
Fig.~\ref{fig:2}(b) and (c). Here in Fig.~\ref{fig:SM1} we present results of 
all non-zero pairing channels. The subleading chaneels evolve in the same way 
as the leading one with the same symmetry.

\section{Cross between the $s_x$ and $s_z$ pairing induced by the shift of 
the $z^2$ bonding band}

\begin{figure}[h!]
	\includegraphics[width=0.5\linewidth]{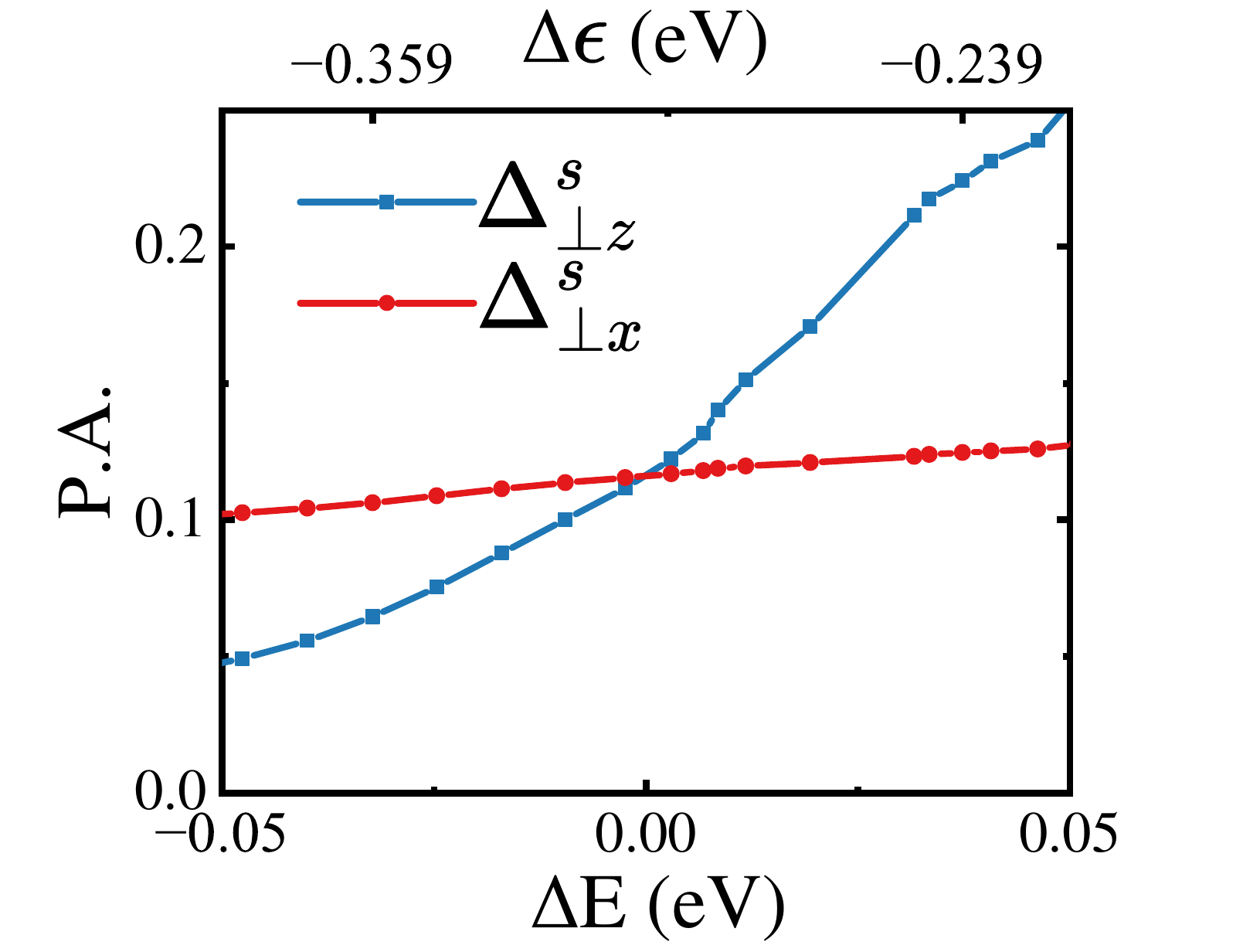}
	\caption{Leading pairing amplitudes (P.A.) as a function of $\Delta E$ and 
	$\Delta \epsilon$ that exhibit a crossover between the $s_x$ and $s_z$ 
	pairing channels.
	}\label{fig:SM2}
\end{figure}

In addition to the transition between the $s_z$ and $d_x$ pairing symmetries 
induced by the shift of the $z^2$ bonding band as discussed in the main text, 
we observe similar pairing cross behavior in other regimes of the phase 
diagram. As shown in Fig.~\ref{fig:SM2}, for $J_{\perp zz}/J_{\parallel 
xx}=2.5$ and $J_{\perp xx}/J_{\parallel xx}=2.5$, the system exhibits a 
crossover between $s_z$ and $s_x$ pairing channels as the band top of the $z^2$ 
bonding band $\Delta E$ is varied. When the $z^2$ bonding band lies above the 
Fermi level ($\Delta E>0$), the $s_z$ pairing dominates, similar to the case 
discussed in the main text. As $\Delta E$ becomes negative and the $z^2$ 
bonding band sinks below the Fermi level, the leading pairing channel of the 
system crosses to be the $s_x$ pairing. This demonstrates that the position of 
the $z^2$ bonding band relative to the Fermi level plays a crucial role in 
determining the exact pairing symmetry across different coupling regimes.

\end{document}